\documentclass[apjl,twocolumn]{emulateapj}
\usepackage{epsfig,mathptmx,graphicx}

\def\gtsima{$\; \buildrel > \over \sim \;$}
\def\ltsima{$\; \buildrel < \over \sim \;$}
\def\prosima{$\; \buildrel \propto \over \sim \;$}
\def\gsim{\lower.5ex\hbox{\gtsima}}
\def\lsim{\lower.5ex\hbox{\ltsima}}
\def\simgt{\lower.5ex\hbox{\gtsima}}
\def\simlt{\lower.5ex\hbox{\ltsima}}
\def\simpr{\lower.5ex\hbox{\prosima}}

\def\h1{$h^{-1}$}
\def\eeq{\end{equation}}
\def\beq{\begin{equation}}

\shorttitle{MAMBO observations of $BzK$ starburst galaxies}
\shortauthors{H. Dannerbauer et al.}

\begin{document}


\title{MAMBO 1.2~ Millimeter observations of B$\lowercase{\rm z}$K-selected star-forming Galaxies at $\lowercase{\rm z}\sim2$\altaffilmark{1}}
\altaffiltext{1}{Based on observations carried out with the IRAM 30~m Telescope at Pico Veleta. IRAM is supported by INSU/CNRS (France), MPG (Germany) and IGN (Spain). Also based on ESO observations (program IDs 072.A-0506, 075.A-0439) and on Subaru observations (program S02B-101, S04A-081), and 
on observations made with the Spitzer Space Telescope, which is operated by the JPL, Caltech under a contract with NASA.}

\author{H.~Dannerbauer\altaffilmark{2},
	E.~Daddi\altaffilmark{3},
	M. ~Onodera\altaffilmark{4,5}
	X.~Kong\altaffilmark{4,6},
	H.~R\"ottgering\altaffilmark{7},
	N.~Arimoto\altaffilmark{4},
	M.~Brusa\altaffilmark{8},
	A.~Cimatti\altaffilmark{9},
	J.~Kurk\altaffilmark{2,9},
	M.D.~Lehnert\altaffilmark{8},
	M. ~Mignoli\altaffilmark{10},
	A.~Renzini\altaffilmark{11}
}

\altaffiltext{2}{MPIA, K\"onigstuhl 17, D-69117 Heidelberg, Germany; dannerb@mpia-hd.mpg.de}
\altaffiltext{3}{{\em Spitzer} Fellow; NOAO,
950 N. Cherry Ave., Tucson, AZ, 85719} 
\altaffiltext{4}{NAOJ, Mitaka, Tokyo 181-8588, Japan}
\altaffiltext{5}{Department of Astronomy, University of Tokyo, Tokyo 113-0033, Japan}
\altaffiltext{6}{CfA, Univ. of Science and 
Technology of China, Hefei 230026, China}
\altaffiltext{7}{Sterrewacht Leiden, PO Box 9513, 2300 RA, Leiden, Netherlands}
\altaffiltext{8}{MPE, Postfach 1312, D-85741 Garching, Germany}
\altaffiltext{9}{INAF-Arcetri, Largo E. Fermi 5, I-50125 Firenze, Italy}
\altaffiltext{10}{INAF-Bologna, via Ranzani 1, I-40127 Bologna, Italy }
\altaffiltext{11}{INAF-Padova, Vicolo dell'Osservatorio 5,
I-35122 Padova, Italy}



\begin{abstract}
We present MAMBO 1.2~mm observations of five $BzK$-pre-selected
vigorous starburst galaxies at z~$\sim2$.  Two of these were detected
at more than 99.5\% confidence levels, with 1.2~mm fluxes around
1.5~mJy.  These millimeter fluxes imply vigorous activity with
star-formation rates (SFRs) $\approx$500--1500$~M_{\sun}~yr^{-1}$,
confirmed also by detections at 24$\mu$m with the MIPS camera on board
of the Spitzer satellite.  The two detected galaxies are the ones in
the sample with the highest SFRs estimated from the rest-frame UV, and
their far-IR- and UV-derived SFRs agree reasonably well. This is
different from local ULIRGs and high-z submm/mm selected galaxies for
which the UV is reported to underestimate SFRs by factors of 10--100,
but similar to the average $BzK$--ULIRG galaxy at $z\sim2$.  The two
galaxies detected at 1.2~mm are brighter in $K$ than the typical
NIR-counterparts of MAMBO and SCUBA sources, implying also a
significantly different $K$-band to submm/mm flux ratio. This suggests
a scenario in which $z\sim2$ galaxies, after their rapid (sub)mm
brightest phase opaque to optical/UV light, evolve into a longer
lasting phase of $K$-band bright and massive objects.  Targeting the
most UV active $BzKs$ could yield substantial detection rates at
submm/mm wavelengths.
\end{abstract}





\section{Introduction}

In the last few years, selection techniques at different wavelengths
have revealed a variety of, at first glance, different high redshift
source populations (e.g., Steidel et al. 1996; 2004; Smail et
al. 1997; Kurk et al. 2000; Franx et al. 2003; Daddi et al. 2004a;
Cimatti et al. 2004).  Highly debated questions are how these objects
are linked to each other; how much they overlap with each other; and
what the evolutionary path is, if any, between them and with the
different types of low redshift galaxies.  Observations in the local
universe are indicating \citep[e.g.,][]{tho05,nel05} that the stars in
massive early-type galaxies formed in general at high redshift,
$z>>1$, with short formation timescales, suggesting high peak
star-formation rates (SFRs). Populations of high redshift galaxies,
among those mentioned above, having large stellar masses and high SFRs
have indeed also been found. However a detailed characterization of
the formation of early-type galaxies is still missing.

Recently, \citet{dad04b} presented a technique based on
optical/near-IR photometry in the $B$, $z$ and $K$ bands that, at
least to $K_{Vega}<20$, allows to obtain virtually complete samples of
galaxies at redshift $1.4<z<2.5$, including star-forming and passively
evolving galaxies. This technique is reddening independent for
$z\sim2$ star-forming galaxies, that in $K<20$ samples have SFRs
estimated between $50-1000$ solar masses per year. These SFRs require
large and possibly uncertain reddening corrections.  The average
long-wavelength emission properties of $BzK$ star-forming galaxies
($sBzK$s hereafter) at $z=2$ have been measured accurately by
\citet{dad05}, finding that the typical $sBzK$ is an Ultra Luminous
Infra-Red Galaxy (ULIRG) with $SFR\sim200-300~M_{\sun}~yr^{-1}$.  As
the $BzK$ technique is substantially complete for $z\sim2$ galaxies,
one expect that also submm/mm selected galaxies (SMGs, see Blain et
al. 2002 for a review) should fulfill the $BzK$ criteria, as many of
them lie at $z\sim2$ \citep[e.g.,][]{cha05}, and that the most active
$sBzK$ galaxies could actually be SMGs.  In this letter we present
observations with the 117--element {\it Max-Planck Millimeter
Bolometer Array} \citep[MAMBO;][]{kre98} of five luminous
($K_{Vega}\simlt19$) $sBzK$ galaxies, with high expected SFRs, based
on their optical/UV properties.  We report the detection of two of the
five galaxies, and discuss some implications on the relation between
$sBzK$ and SMGs, and on the assembly process of massive galaxies at
high redshifts.  We assume a Salpeter initial mass function (IMF) from
0.1 to 100 $M_\odot$, and a cosmology with $\Omega_\Lambda, \Omega_M =
0.73, 0.27$, and $h = H_0$[km s$^{-1}$ Mpc$^{-1}$]$/100=0.71$
\citep{spe03}.

\section{Sample selection}
\begin{center}
\begin{deluxetable*}{lcccccccccccl}
\setlength{\tabcolsep}{0.02in}
\tablecaption{Properties of the $sBzK$ galaxies observed with MAMBO.\label{tab:bzkprop}}
\tablewidth{0pt}
\tablehead{
\colhead{Source} & \colhead{RA} & \colhead{DEC} & \colhead{$K$} & \colhead{$BzK$} &
\colhead{$z_{spec}$} & \colhead{E$(B-V)$} &\colhead{SFR$_{UV}$} & \colhead{Mass} & \colhead{S$_{1.2~mm}$} & \colhead{SFR$_{IR}$} & \colhead{S$_{24~\mu m}$} & \colhead{Comments}\\
\colhead{} & \colhead{(J2000)} & \colhead{(J2000)} & \colhead{(AB)} & \colhead{(AB)} &
\colhead{} & \colhead{} & \colhead{($M_{\sun}/yr$)} & \colhead{($10^{11}M_{\sun}$)} & \colhead{(mJy)} & \colhead{($M\sun/yr$)} & \colhead{($\mu$Jy)} & \colhead{}\\
\colhead{(1)} & \colhead{(2)} & \colhead{(3)} & \colhead{(4)} &
\colhead{(5)} & \colhead{(6)} & \colhead{(7)} &\colhead{(8)} & 
\colhead{(9)} & \colhead{(10)} & \colhead{(11)} & \colhead{(12)} &
\colhead{(13)}
}
\startdata
OBJ4415 & 14:48:35.35 & 08:52:27.0 & 20.92&1.61&2.53&0.49&177/140&2.8&0.55$\pm$0.54&&&Ly$\alpha$; AGN-2\\
OBJ4193$^a$ & 14:48:33.23 & 08:54:14.1 & 20.67&1.99&2.76&0.40&$\sim$200&4.0& --0.26$\pm$0.58&&&Ly$\alpha$; AGN-2\\
OBJ2742 & 14:49:20.50 & 08:50:52.3 & 20.84&0.02&2.16$^{b}$&0.53&666/400&1.4&1.34$\pm$0.43$^{c}$&700--1400&70$\pm$23&UV-abs/IR-em?\\
OBJ2426 & 14:49:28.96 & 08:51:52.9 & 20.76&0.05&2.37&0.59&546/1400&1.8&1.50$\pm$0.42&800--1900&200$\pm$50&UV-abs/IR-em\\
OBJ1901 & 14:49:41.38 & 08:59:50.5 & 20.58&0.68&1.60&0.33&186/140&1.7&0.54$\pm$0.43&&$79\pm25$&UV-abs/IR-em\\
\enddata
\tablecomments{ 
Col.~(1): Object identification.  Cols.~(2)-(3): Optical/NIR coordinates. 
Col.~(4): Total $K$-band magnitudes.
Col.~(5): $BzK$-color. $BzK\equiv(z-K)_{AB}-(B-z)_{AB}$ \citep{dad04b}. 
Col.~(6): Spectroscopic redshift. Col.~(7): Colour excess: $E(B-V)=0.25[(B-z)_{AB}+0.1]$, Daddi et al (2004b). Col.~(8): Two SFR
estimates are from the UV corrected 1500~\AA\ luminosities \citep{dad04b} 
and  from SED fitting, respectively. 
Col.~(9): Stellar mass estimated from SED fitting; uncertainties are typically
a factor of two.  Col.~(10): MAMBO ``on--off'' flux at 1.2~mm. Col.~(11):
SFRs estimate based on $L_{IR}$. Col.~(12): MIPS 24~$\mu$m flux. Col.~(13): 
Comments to the UV/optical spectra of individual sources. 
$^{a}$: OBJ4193 is at the edge of the Subaru imaging data, photometry is less accurate.
$^{b}$: this redshift is lower quality, although supported by
UV-spectrum and possible H$\alpha$ detection.
$^{c}$: Excluding the first observing consisting of 7~scans under mediocre 
weather conditions would give $S_{1.2~mm}=1.57\pm0.46$~mJy ($\tau_{1.2~mm}\leq0.2$).}
\end{deluxetable*}
\end{center}

We selected for MAMBO observations five bright $K_{Vega}<19.2$~mag
$sBzK$-galaxies (see Table~\ref{tab:bzkprop}), taken from the 700
arcmin$^2$ wide 'Daddi-Field' at 14h \citep{dad00,kon05}.  All the
targets had spectroscopic redshift identifications from deep VLT-VIMOS
optical spectroscopy and sometimes VLT-SINFONI and/or Subaru-OHS/CISCO
near-IR spectroscopy, as a part of our ongoing survey for $BzK$
selected galaxies \citep{kon05}.  The optical and near-IR
spectroscopic observations will be presented elsewhere. Redshifts
range from 1.6 to 2.8, and SFRs estimated from the dust corrected UV
luminosity range from $\sim150$ to $\sim1400~M_{\sun}~yr^{-1}$.
Although the K-band brightest $sBzK$ naturally tend to have fairly red
average optical/UV colors, preference among sources with measured
redshifts was given to objects with red $B-z$ colors, corresponding to
large $E(B-V)$. The reasons for this preference are twofold: 1)
optical redness might be an evidence for the presence of large amounts
of dust, enhancing the chance of detections with MAMBO; 2) the red
colors ensure these objects would rarely qualify for the $z=2$
criteria of \citet{stei04}, thus allowing to investigate to which
extent vigorous starburst galaxies may be missed by surveys selected
in the UV.

The field has been observed in the X-ray with XMM-Newton for 80 ks
\citep{bru05} and with Chandra as a mosaic of $3\times30$~ks with an
overlapping area of 90~ks (Brusa et al. in preparation).  Targets for
MAMBO observations were chosen to be un-detected in the X-rays
(corresponding to X-ray luminosities $\simlt10^{43}$ erg s$^{-1}$) and
with no strong AGN feature in their UV/optical spectra. However, two
of the targets (OBJ4415 and 4193) show faint and narrow type2 AGN
emission lines in their (almost continuum-less) UV spectra, similar to
some of the objects described by \citet{dok03}. These two objects were
not detected at 1.2~mm.

\section{Observations and results}

Each of the five $sBzKs$ were observed under predominantly excellent
weather on several days during December 2004 with MAMBO--117, at the
IRAM 30~m telescope on Pico Veleta. MAMBO operates at an effective
frequency of 250~GHz, corresponding to 1.2~mm
(FWHM~$\approx10.7$~\arcsec).  The science targets were observed in
``on-off'' observing mode, which is the photometric mode of MAMBO, for
2700 to 5700~s. This observing technique is based on the chop-nod
technique where the target is placed on a reference bolometer element
(on-target channel).  A detailed description of this now standard
observing technique can be found in e.g., \cite{ber00,lut05}.  In
order to reduce systematic errors and avoid spurious detections, the
targets were observed several times and on different days.  The data
were reduced using MOPSIC\footnote{
http://www.astro.ruhr-uni-bochum.de/nielbock/simba/mopsic.pdf}, an
upgraded version of MOPSI \citep{zyl98}.  Every scan with its
individual subscans were carefully inspected for the presence of
possible outliers or anomalies or influence of high opacity
($\tau_{1.2~mm}>0.2-0.3$). From each channel we subtracted the
correlated skynoise from the surrounding channels.  The outstanding
weather conditions -- opacity below $\tau_{1.2~mm}<0.2$ during 60\% of
the time -- ensured stability and high data quality over the several
days in which the data where gathered.

\begin{figure}
\begin{center}
\includegraphics[width=8.1cm]{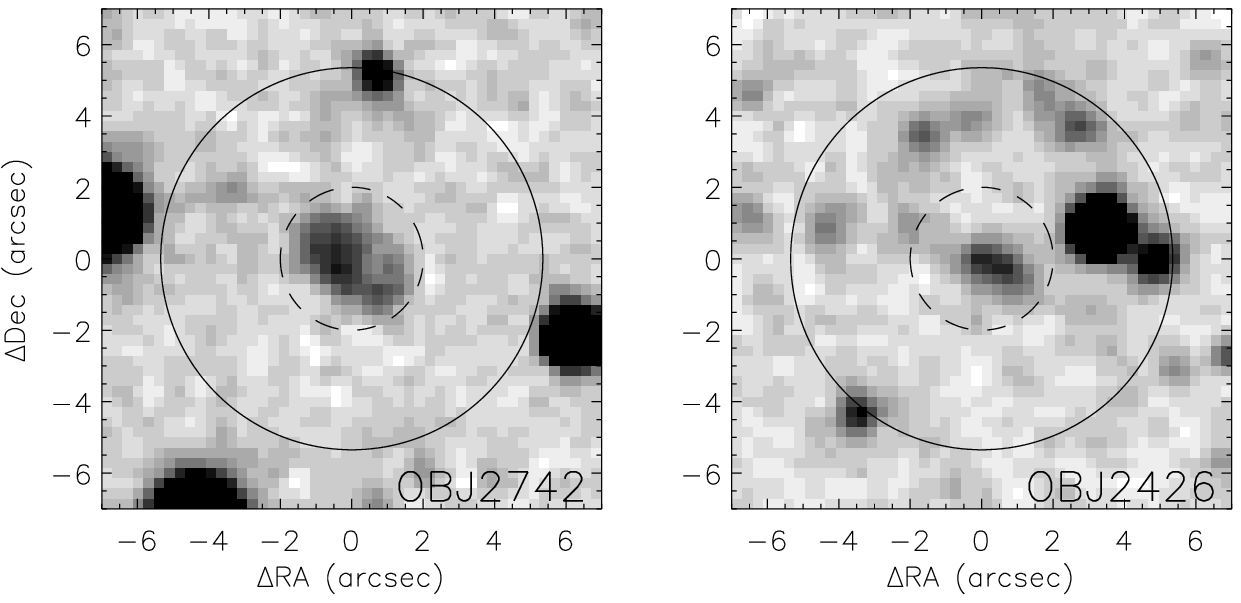}\\
\includegraphics[width=8.1cm]{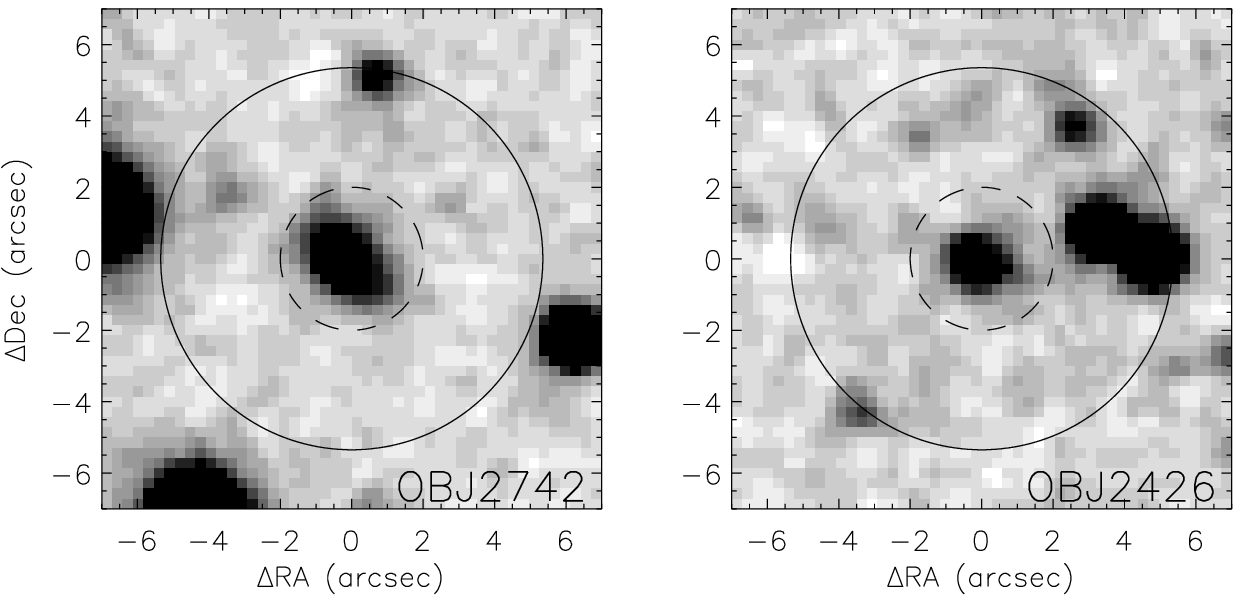}
\caption{B-band (top) and z-band (bottom) images for the fields around
the two MAMBO detections (dashed circle): OBJ2742 and OBJ2426. The
beam size of MAMBO (10.7\arcsec~FWHM, big circle) is fully contained
in the images.  Both sources show an irregular, possibly merger-like morphology. North is up and east to the left.}
\label{fig:mambob}
\end{center}
\end{figure}

A flux of about 1.3--1.5~mJy (significant at $3.2-3.6\sigma$~level) is
detected in the reference bolometers corresponding to the positions of
two of the five $sBzK$ galaxies (OBJ2742 and OBJ2426; see
Table~\ref{tab:bzkprop}).  The MAMBO array of detectors consist of 117
bolometers spread over a sky-region of about 4 arcmin in diameter. In
order to establish reliably the significance of the detections, and to
estimate the likelihood that these may be just be obtained by chance,
we independently reduced the observations for each of the 117 MAMBO
bolometers.  Considering the 5 independent objects observed, this
provides a large control sample of about 550 reliable bolometers.
Given the source counts at 1.2~mm \citep{gre04}, the chance
probability to observe a galaxy with flux $\simgt1.5$mJy within half a
beam FWHM is 0.6\%.  However, we assume conservatively that all the
control bolometers are seeing a blank mm sky.  No control bolometer
has a detection with $S/N>3.6$ (OBJ2426), while 3 control bolometers
have $S/N>3.2$ (OBJ2742). Therefore we conclude that the detections
are reliable at better than the 99.5\% confidence level.
Fig.~\ref{fig:mambob} shows beam sized fields in the B and z band
around the two detections.

In order to constrain better the IR-SED of both MAMBO detections, and
to further improve the confidence in the detections, we obtained MIPS
24$\mu$m imaging data of the field from the Spitzer archive.  In the
500s per sky-pixel data, OBJ2742 is detected at a 3~$\sigma$ level
with a flux $S_{24~\mu m}\approx70~\mu Jy$.  OBJ2426 is brighter with
a flux $S_{24~\mu m}\approx200\pm50~\mu Jy$ (the large error being due
to the fact that this object is blended with a nearby source 3.5$''$
to the west, see Fig.~\ref{fig:mambob}, that has a similar 24$\mu$m
luminosity).  The $24~\mu m$/$1.2~mm$ flux ratios for our two $sBzKs$
are fully consistent with the range reported by \citet{ivi04} for
MAMBO galaxies, and also consistent with SCUBA sources
\citep{fray04,ega04}.

The two galaxies detected are the ones with the highest estimated SFRs
based on the rest-frame UV. No other clearcut differences e.g., color,
magnitude, could be found between the MAMBO detected and undetected
$sBzK$ galaxies.  A posteriori, it is reasonable and reassuring that
we detect with MAMBO the 2 sources that were estimated already to be
the most active in our sample.  This further enhance the confidence
that the detections are reliable.

\begin{figure}[!h]
\begin{center}
\includegraphics[width=8.1cm]{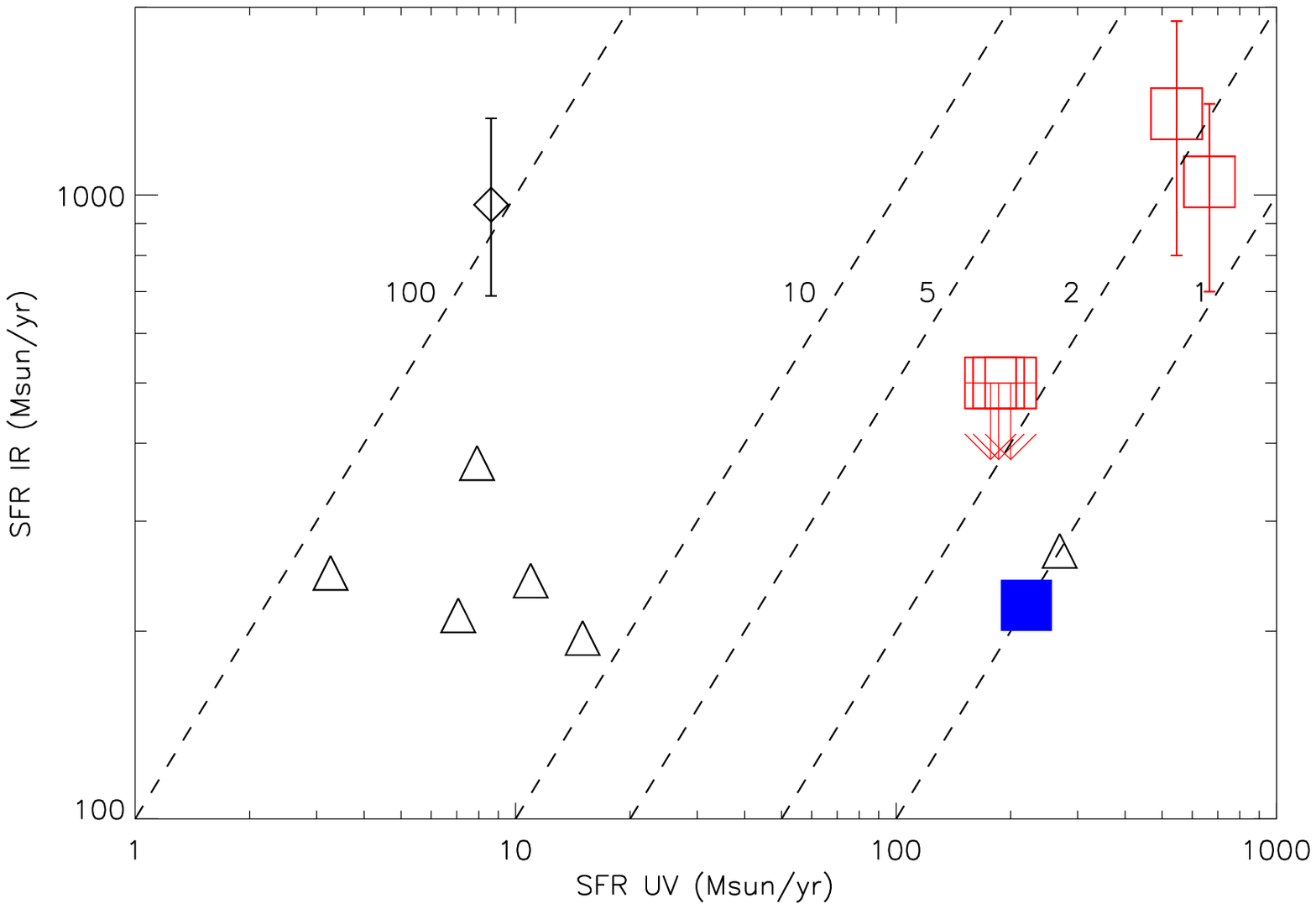}
\includegraphics[width=8.1cm]{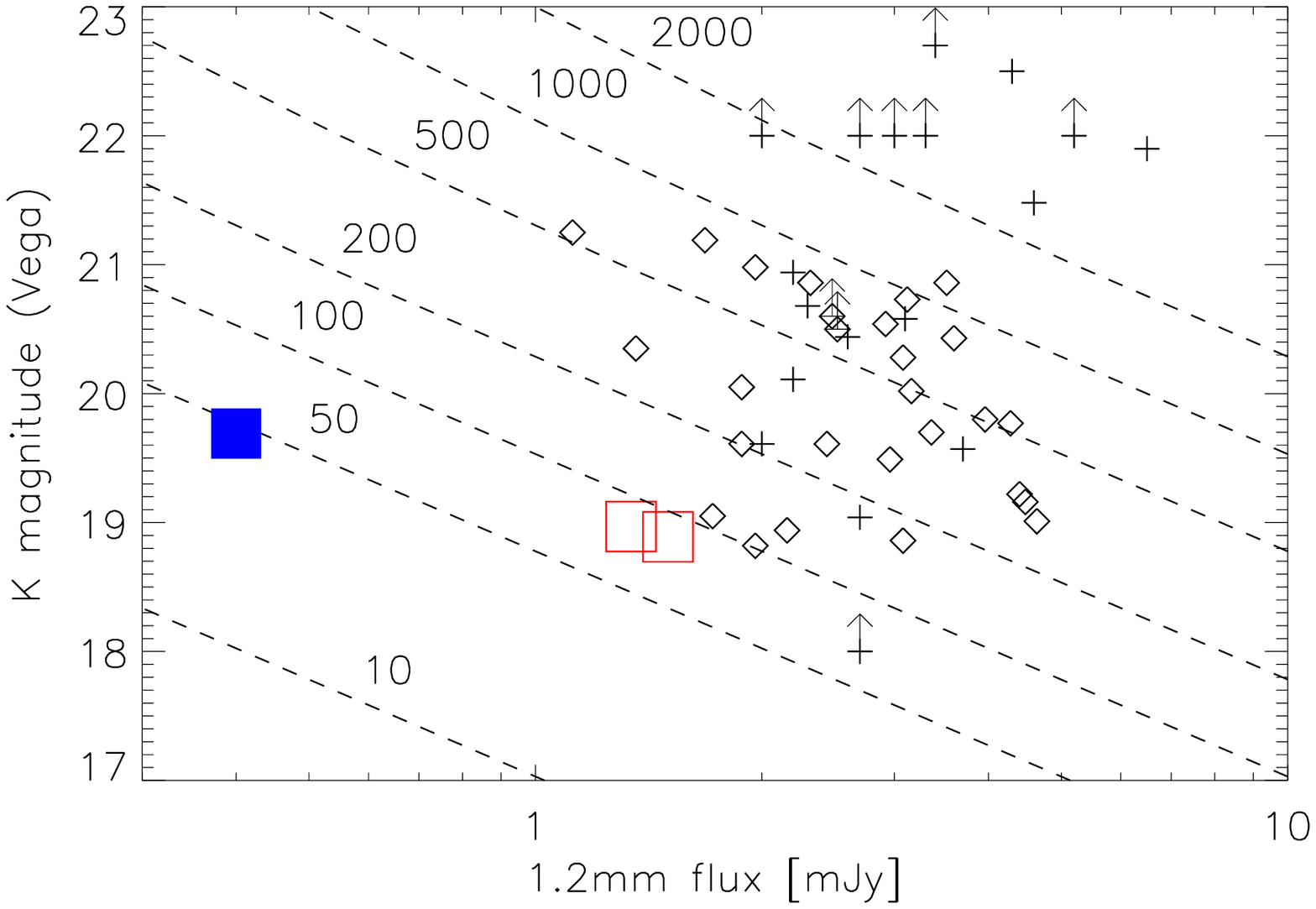}
\caption{{\it(Top Panel)} SFR$_{IR}$ vs. SFR$_{UV}$ 
 for the five $sBzKs$ observed by MAMBO (red open
squares). For comparison, we show the median SFR of $sBzKs$ from the
GOODS-N field \citep[blue filled square;][]{dad05}, a sample of local ULIRGs \citep[open triangles;][]{gol02,tre99}, and the median for
radio-identified SMGs from the SCUBA redshift survey \citep[open
diamond;][]{sma04,cha05}. The dashed lines represent different ratios
of SFR$_{IR}$ to SFR$_{UV}$ from 1 to 100.  {\it(Bottom Panel)} K-band
magnitude (Vega) vs. 1.2~mm flux for our two $sBzKs$ detected by
MAMBO (red open squares). For comparison, we show the median of $sBzKs$ from the GOODS-N
field (blue filled square), radio-identified MAMBO galaxies from the NTT Deep
Field \citep[crosses;][]{dan04}, and SMGs with $z\ge1.5$ from the SCUBA
redshift survey (open diamonds). The dashed lines represent different ratios
of 1.2~mm to K-band flux from 10 to 2000.}
\label{fig:sfr}
\end{center}
\end{figure}

\section{Discussion}

The fluxes measured at 1.2mm (close to the IR-peak of the dust
emission) and at 24$\mu$m with Spitzer (in the mid-IR rest-frame
region) imply that the two detected $sBzKs$ are active IR galaxies.
To obtain accurate estimates of their total IR luminosities
(8--1000$\mu$m) would require further observations between 24$\mu$m
and 1.2mm, and especially in the range 60--100$\mu$m rest-frame, that
could constrain the temperature of the dust. For $z=2$, obtaining
these data to the required depths is beyond the possibilities of the
available observing capabilities. However, we can derive constraints
on the total IR luminosities (and thus SFRs) by assuming these 2
detections have IR SEDs similar to those of well studied $z=2$
populations, i.e. the average $sBzK$ galaxies (Daddi et al. 2005) or
SCUBA selected galaxies (Chapman et al. 2005), or by assuming that
local empirical correlations between mid-IR or submm luminosities and
$L_{IR}$ (e.g. Chary \& Elbaz 2001) holds also at $z=2$. We find that
this suggests IR luminosities in the range of
$L_{IR}\approx3$--$10\times10^{12}L_\odot$, which correspond to SFRs
in the range 500--1500$M_{\sun}~yr^{-1}$, the factor of 3 uncertainty
being indicative of the range of estimates from the different methods.

We considered the possibility that the detected mm fluxes are due to
AGN emission. The submm-to-X-ray spectral slope $\alpha_{SX}$
(Alexander et al. 2003) is an indicator of AGN activity.  For the two
MAMBO-detected $sBzK$s we derive an upper limit of
$\alpha_{SX}\simlt1.20$\footnote{We assume ratios of $2.5-3$ between
850$\mu$m and 1.2~mm fluxes, that are the factors spanned by the
models at the luminosities of our objects.} that, if compared to
Fig.~7 in \citet{ale03}, points toward the starburst part of the
diagram occupied, e.g., by SCUBA galaxies in HDFN and Arp~220, and
exclude both luminous unobscured (3C273) and obscured quasars. Of
course, this does not allow us to completely discard the presence of
Compton thick AGNs (NGC~6240).  However, the lack of any AGN features
in the UV spectra, together with the reasonable agreement between UV-
and far-IR- estimated SFRs, suggest that the far-IR activity of these
sources is powered by star formation.

An interesting result of our observations is indeed that the IR
luminosities and related SFRs inferred for our objects from the mm
fluxes are in quite a good agreement, at most a factor 2--3 larger
than those estimated from the dust-reddening corrected UV luminosities
(Fig.~\ref{fig:sfr}, top panel). This agreement, already noticed by
\citet{dad05} for the typical $sBzK$ galaxy, is somewhat surprising. A
fairly different situation seemingly holds for submm/mm selected
galaxies (Chapman et al. 2005; Swinbank et al. 2004) and for local
far-IR selected ULIRGs (Goldader et al. 2002, with only a notable
exception), where the observed IR luminosities and far-IR estimated
SFRs exceed by 1--2 orders of magnitude the same quantities estimated
from the dust corrected UV luminosities (Fig.~\ref{fig:sfr}, top
panel).  Extending this comparison further (Fig.~\ref{fig:sfr}, bottom
panel), we notice that $sBzKs$ galaxies in this paper and in Daddi et
al. (2005) are generally brighter in the $K$ band ($\sim18-20$~mag)
than the typical counterparts of MAMBO galaxies
\citep{ber00,dan02,dan04} and SCUBA sources
\citep[e.g.,][]{bla02,sma04}, although they lie at a similar redshift
$z\sim2$. The mm/submm fluxes of SMGs are instead generally higher
than those of $sBzKs$ (Fig.~\ref{fig:sfr}).  This implies a
significantly different $K$-band to submm/mm flux ratio between far-IR
selected vs. optically selected $z\sim2$ galaxies.

Similarities also exist between $sBzK$s and far-IR selected
populations both in their irregular morphologies (Conselice et al.
2004; Trentham et al. 1999; Daddi et al. 2004a; this paper, see
Fig.~\ref{fig:mambob}) and in their red observed UV slopes (i.e. high
apparent reddening; Chapman et al. 2005; Daddi et al. 2004b; 2005;
Goldader et al. 2002; this paper, see Table~1).  The differences
(submm/mm flux levels, $K$-band to submm/mm ratios and $SFR_{UV}$ to
$SFR_{IR}$ ratios) and similarities (morphologies, UV slopes) are most
likely effects induced by the different selection of the samples
($K$-band versus far-IR), and can be reasonably understood by assuming
that the far-IR emitting region of most SMGs is observed behind a
thick dust screen, nearly opaque to the optical and near-IR radiation.
This can be fit into an evolutionary scenario in which the starburst
galaxies are initially deeply embedded into a large column density of
dust, during the submm/mm brightest phase when substantial obscuration
of a few magnitudes or more is present even at observed $K$-band
(rest-frame $5000-8000$~\AA\ for $z=2-3$). The UV light is almost
completely attenuated behind the dust screen so that it is impossible
to measure the ongoing SFR from the residual UV light transmitted. As
time passes by, SFR and L$_{IR}$ are reduced and the dust is diffused
and ejected out of the main star-formation regions (presumably by
galactic winds or other kinds of feedback), or simply destroyed. Thus
the resulting galaxy is brighter in the near-IR and transparent enough
in the rest frame UV that the transmitted light can account for the
ongoing SFR after reddening correction.  The submm brightest phase has
presumably a short duration \citep[$\sim40-200~Myr$;][]{sma04,gre05},
while the subsequent {\em normal} $sBzK$ phase is estimated to last
quite longer ($\sim0.5$--1~Gyr; Daddi et al. 2005), which would
explain why SMGs have a factor of 10 or more lower spatial density
\citep{sco02,dad04a,dad04b}, while having similar masses
\citep[e.g.,][]{gen03,gre05} and perhaps spatial clustering
\citep{bla04,dad04a,ade05}.  This picture is consistent with the
evolutionary scenario proposed by Vega et al. (2005).  Chapman et al.
(2005) also suggest that SMGs evolve into near-IR bright galaxies.

We conclude by suggesting that the follow-up of the most UV active
$sBzK$ could be a promising alternative for finding submm/mm sources,
possible complementary to the radio pre-selection method presented by
\citet{cha01} for a similar redshift range $z\sim1.5-2.8$, or the
Spitzer$+$MIPS pre-selection (Lutz et al 2005).  \smallskip

We would like to thank the IRAM staff at Pico Veleta for their support
of these observations; Fabian Walter for his careful reading of an
earlier version of this manuscript and his insightful and helpful
comments; Axel Wei\ss\ for useful conversation about the MAMBO data;
and Ian Smail for useful discussions.  ED gratefully acknowledges NASA
support through the Spitzer Fellowship Program, under award
1268429. This work is partly supported by a Grant-in-Aid for
Scientific Research (No. 16540223) by the Japanese Ministry of
Education, Culture, Sports, Science and Technology.

\clearpage

\end{document}